\documentclass[prb,twocolumn,showpacs,amsfonts,amssymb,floats,superscriptaddress,aps]{revtex4}

\usepackage{amssymb}
\usepackage{graphicx}
\usepackage{amsmath,color}

\newcommand{\s}{\sigma}
\newcommand{\e}{\epsilon}

\newcommand{\klammer}[3]{\left#1 #3 \right#2 }
\newcommand{\klammerB}[3]{\Big#1 #3 \Big#2 }
\newcommand{\klammerb}[3]{\big#1 #3 \big#2 }
\newcommand{\bra}[1]{\langle #1|}

\newcommand{\ket}[1]{| #1\rangle}

\newcommand{\ew}[1]{\langle #1\rangle}
\newcommand{\comment}[1]{}
\begin{document}

\title{Dynamic susceptibilities of the single impurity Anderson model within an
enhanced non-crossing approximation}

\author{Sebastian Schmitt}
\affiliation{Institut f\"{u}r Festk\"{o}rperphysik, Technische Universit\"{a}t Darmstadt,
Hochschulstr. 6, D-64289 Darmstadt, Germany}
\affiliation{Lehrstuhl f\"{u}r Theoretische Physik II, Technische Universit\"{a}t Dortmund,
Otto-Hahn Str. 4, D-44221 Dortmund, Germany}
\author{Torben Jabben}
\affiliation{Institut f\"{u}r Festk\"{o}rperphysik, Technische Universit\"{a}t Darmstadt,
Hochschulstr. 6, D-64289 Darmstadt, Germany}
\author{Norbert Grewe}
\affiliation{Institut f\"{u}r Festk\"{o}rperphysik, Technische Universit\"{a}t Darmstadt,
Hochschulstr. 6, D-64289 Darmstadt, Germany}

\date{\today}

\begin{abstract}
The single impurity Anderson model (SIAM) is studied within an   enhanced 
non-crossing approximation (ENCA). This method is extended to the
calculation of susceptibilities and thoroughly tested, also in order
to prepare applications as a building block for the calculation of
susceptibilities and phase transitions in correlated lattice
systems. A wide range of model parameters, such as 
impurity occupancy, temperature, local Coulomb repulsion and hybridization strength,
are studied.
Results for the spin and charge susceptibilities are presented.
By comparing the static quantities to exact Bethe ansatz results, it is shown
that the description of the magnetic excitations of the impurity within the ENCA 
is excellent, even in situations with large valence fluctuations or vanishing Coulomb repulsion.
The description of the charge susceptibility 
is quite accurate in situations where the singly occupied ionic configuration is the
unperturbed ground state; however, it seems to overestimate charge fluctuations in the asymmetric 
model at too low temperatures.
The dynamic spin excitation spectrum is dominated by the Kondo-screening of the 
impurity spin through the conduction band, i.e.\ the formation of the local Kondo-singlet.
A finite local Coulomb interaction $U$ leads to a drastic reduction of the charge response 
as processes involving the doubly occupied impurity state are suppressed. 
In the asymmetric model,
the charge susceptibility is enhanced for excitation energies smaller than 
the Kondo scale $T_K$ due to the influence of valence fluctuations.  
\end{abstract}


\pacs{75.40.Gb,71.27.+a, 72.10.Fk}

\maketitle

\section{\label{sec:introduction} Introduction}

The single impurity Anderson model (SIAM)
describes an impurity of localized $f$-states 
with local Coulomb interaction embedded into a
metallic host of non-interacting $c$-band electrons.\cite{anderson:siam61}
In its simplest version it discards the possibility of a complex orbital 
structure of the impurity and models the local $f$-states through a two-fold 
degenerate $s$-orbital. The Hamiltonian for the impurity  reads
\begin{align}
  \label{eq:siamHamF}
  \hat{H}_f&=\sum_\s \left(
    \e^f \:\hat{f}^\dagger_{\s}\hat{f}_{\s}+\frac{U}{2}\, \hat{n}^f_{\s}\hat{n}^f_{\bar{\s}}
  \right)  \qquad,
\end{align}
with $\hat{f}^\dagger_{\s}$ ($\hat{f}_{\s}$)  and  $\hat{n}^f_{\s}=\hat{f}^\dagger_{\s}\hat{f}_{\s}$
the usual creation (annihilation) 
and  number operators for $f$-electron with spin $\s$, respectively.
The local one-particle energy
is given by $\e^f$, and the local Coulomb interaction is
the usual density-density interaction proportional to the matrix element $U$.
The non-interacting conduction electrons  are modeled by a single band of Bloch states
with crystal momentum $\underline{k}$
characterized by the dispersion relation $\e^c_{\underline{k}}$,
\begin{align}
  \label{eq:siamHamC}
  \hat{H}_c=\sum_{\underline{k},\s}\e^c_{\underline{k}}\,\hat{c}^\dagger_{\underline{k}\s}\hat{c}_{\underline{k}\s}
  \quad.
\end{align}
These two parts mix via a hybridization amplitude $V_{\underline{k}}$,
\begin{align}
  \label{eq:siamHamV}
  \hat{V}&=\frac{1}{\sqrt{N_0}}\sum_{\underline{k}} \left(V_{\underline{k}}\,\hat{f}^\dagger_{\s}\hat{c}_{\underline{k}\s}+h.c. \right) 
  \quad.
\end{align}
The total Hamiltonian is then the sum of these three terms 
\begin{align}
  \label{eq:app-siamHam}
  \hat{H}&=\hat{H}_c+\hat{H}_f+\hat{V}
  \quad.
\end{align}

Even though the thermodynamics of the model can be solved 
exactly within the Bethe ansatz  
method,\cite{andreiSolutionKondo83,wiegmannExactSIAMI83,tsvelickExactSIAMII83,tsvelick:ExactSIAM83}
dynamic quantities can  in general not be obtained
exactly and one has to rely on approximations. 
The SIAM has been extensively studied with various methods, including 
the numerical renormalization group (NRG),\cite{wilsonNRG75,krishnamurtyNRGSIAMI80,krishnamurtyNRGSIAMII80,bullaNRGReview08}
the (dynamic) density-matrix renormalization group ((D-)DMRG),\cite{whiteDMRG92,whiteDMRG93,kuehnerDDMRG99,hallbergDMRGreview06}
quantum Monte Carlo  (QMC) methods \cite{hirschQMC86,rubtsovContTimeQMC05,wernerContinousTimeQMC06}
and direct perturbation theory with respect to the hybridization.\cite{keiter:PerturbJAP71,keiter:PerturbIJM71,grewe:IVperturb81,keiterMorandiResolventPerturb84}
Especially with the development of the dynamical mean-field theory 
(DMFT),\cite{georges:dmft96} where the solution of an effective SIAM represents 
the essential step towards the solution of the correlated lattice system,  the interest 
in accurate and manageable impurity solvers
has increased. 

In this work, we extend the well established 
enhanced non-crossing approximation
(ENCA)\cite{pruschkeENCA89,greweCA108,holmFiniteUNCA89,keiterENCA90} 
to the calculation of the static and dynamic 
susceptibilities of the impurity. Like many other 
approximations formulated within the direct perturbation
theory with respect to the hybridization,\cite{Grewe:siam83,Kuramoto:ncaII84,sakaiCoreLevel88,
  anders:PostNCA95,kroha:CTMA97,haule:sunca01,greweCA108}
the ENCA is thermodynamically conserving in the sense of Kadanoff and 
Baym.\cite{baymKadanoffConservation61,baym:conservingApprx62}
It extends the usual non-crossing approximation (NCA) to finite values of 
the Coulomb repulsion $U$ via the 
incorporation of the lowest order vertex corrections, which
are necessary to produce the correct Schrieffer-Wolff exchange coupling
and the order of magnitude of the low energy Kondo scale of the problem.
From the NCA it is well known that some pathological structure appears at the
Fermi level below a
pathology scale\cite{kuramoto:AnalyticsNCA85,bickers:nca87a} $T_{path}\approx 10^{-1}-10^{-2}T_K$.
The ENCA removes the cusps in  spectral functions 
associated with this pathology\cite{pruschkeENCA89}
and only a slight overestimation of the height of the many-body resonance
at very low temperatures remains.
As it will be shown in this work, the 
skeleton diagrams selected within the ENCA suffer from an
imbalance between charge and spin excitations
and overestimate the influence of charge fluctuations. 
Other than that, it has no further limitations.

Despite the known limitations of the NCA it has been widely applied to more 
complex situations due to the forthright possibility of  extensions. For 
the SIAM out of equilibrium it is one of the few methods to incorporate
nonequilibrium dynamics as well as many-body effects.\cite{wingreenNonEqNCA94}
Complex orbital multiplets can be included in a straight-forward manner
and connections to experimental data can be made.\cite{Ehm2007} 
However, incorporating finite values of $U$ may change the many-body features near 
the Fermi level considerably.\cite{greweCA108,greweENCACF09} Therfore, a well tested 
extension of the ENCA in order to calculate susceptibilities at finite $U$
with the same accuracy as the spectra is desired.
In particular, the calculation of lattice susceptibilities within DMFT 
\cite{schmitt:sus05,jarrell:symmetricPAM95,schmittPhD08} 
needs a reliable strategy for an effective impurity, and
the treatment of one- and two-particle excitations of the same footing is
of paramount importance.
Calculations of lattice susceptibilities 
and phase transitions of the Hubbard model with the ENCA as the impurity solver
are presented elsewhere.\cite{inprep,schmittPhD08}

Compared to the ``numerically exact'' schemes like the renormalization group
methods (RG), exact diagonalization (ED) 
or  QMC,
the direct perturbation theory has its advantages: (i) The approximations
are free of systematic errors  stemming from the discretization of 
the conduction band
(RG and ED) or imaginary time (QMC). 
The continuum of band states is kept throughout the calculation, and dynamic Green functions
are formulated with continuous energy variables. Thus, there 
are no discretization-artefacts~\cite{zitkoArtefactsNRG09} 
and there is no need for artificial broadening parameters\cite{sakaiSIAMRG89,bulla:MITFTNRG01}
or $z$-averaging.\cite{oliveiraZTrickNRG94,campoZtrickNRG05}
(ii) The coupled integral equations for dynamic quantities, which have to be solved numerically,
are  formulated on the real frequency axis, which renders
the non-trivial numerical analytic continuation of a finite set of
Fourier coefficients\cite{jarrellMEM96} 
or deconvolution\cite{raasSpectralDMRG05} unnecessary. 
The continuous-time quantum Monte Carlo
approach (CTQMC)\cite{wernerContinousTimeQMC06} avoids a systematic Trotter-error (i)
but it is still plagued with the occurrence of a negative sign-problem.\cite{troyerQMC05}

The discretization errors (i) are of special importance for self-consistent calculations
like the DMFT. There, the solution of an impurity model is used to construct a
guess for the Green function of the lattice which is then used to 
yield a new effective ``conduction band'' for the impurity model.
Errors in the treatment of the continuum of band states will be propagated by 
the iterative solution  and may lead to an incorrect distribution of spectral 
weight.

The ENCA can be solved  quite effectively on 
simple desktop computers, and is numerically not very demanding.
Spectral functions 
can be calculated within some seconds to minutes 
while  dynamic susceptibilities may take up to some hours. 
Additionally, it contains no free parameters and no fine-tuning is necessary.
This makes it  especially interesting
for involved lattice calculations.

\section{Theory}
\label{sec:theory} 

\subsection{Direct perturbation theory}
\label{sec:perturb} 

\begin{figure*}[t]
  \includegraphics[scale=1]{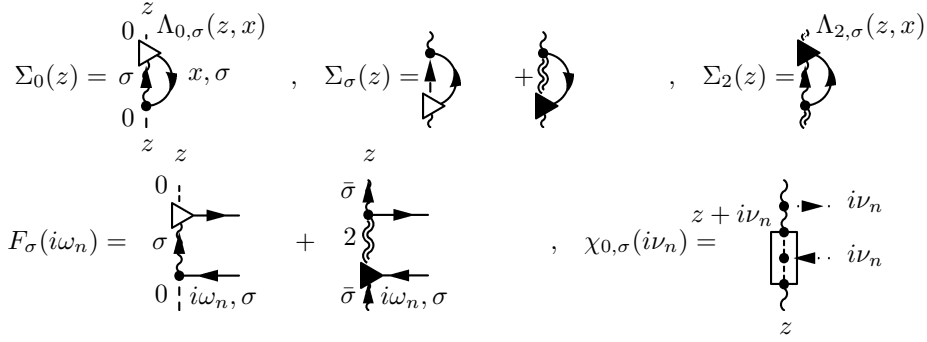}
  \caption{Diagrammatic equations for the ionic self-energies (upper), $f$-electron Green function (lower left) 
    and dynamic susceptibility (lower right, $\chi_{0,\s}(i\nu)$ as an example). 
    The dashed, wiggly and double-wiggly lines represent the empty, singly occupied ($\bar{\s}$ represents 
    the opposite of the spin $\s$) and doubly occupied ionic
    propagators, respectively. The triangles and boxes represent the appropriate vertex functions. 
    The closed full lines in the upper graphs represent uncorrelated propagations in the $c$-band.
  }
  \label{fig:appNCABase}
\end{figure*}

In direct perturbation theory with respect to the hybridization
term $\hat{V}$ the ``unperturbed'' system is represented by the 
uncoupled ($V_{\underline{k}}=0$)  interacting impurity. This is diagonalized 
by the  ionic many-body states $\ket{M}$
\begin{align}
  \label{eq:app-EfDef}
  \hat{H}_f&=\sum_{M}E_M\:\hat{X}_{M,M} 
  \quad,
\end{align}
where the operators $\hat{X}_{M,M}=\ket{M}\bra{M}$ 
are projectors on the eigenstates $\ket{M}$
and are diagonal versions of the so-called 
ionic transfer (or Hubbard)  operators
$\hat{X}_{M,M'}=\ket{M}\bra{M'}$.
For a simple $s$-shell the quantum numbers 
$M$ characterize the  empty $\ket{0}$, 
singly occupied with spin $\s$ $\ket{\s}$
and doubly occupied $\ket{2}$ impurity states
with the corresponding unperturbed  eigenvalues
$E_0=0$, $E_{\s}=\e^f$ and $E_2=2\e^f+U$, respectively.
Furthermore, 
the partition function and dynamic Green function
are expressed in terms of a contour-integration in
the complex plane,
\begin{align}
  \label{eq:app-Znca}
  Z&= Tr\: e^{-\beta\,\hat H}=\oint_\mathcal{C} \!\frac{dz}{2\pi i} 
  \:e^{-\beta\,z}\:Tr\:\klammerb{[}{]}{z-\hat H}^{-1}
  \\
\label{eq:app-Gfnca}
  G_{A,B}(i\eta)&= \frac{1}{Z}\oint_\mathcal{C} \!\frac{dz}{2\pi i} 
  \:e^{-\beta\,z}\:Tr\Big\{\:\klammerb{[}{]}{z-\hat H}^{-1}\:\hat{A}\:\cdot\\
  \notag 
  & \phantom{\frac{1}{Z}\oint_\mathcal{C} \!\frac{dz}{2\pi i}   \:e^{-\beta\,z}\:Tr}
  \cdot\:\klammerb{[}{]}{z+i\eta-\hat H}^{-1} \hat{B}
  \Big\}  \quad,
\end{align}
with $i\eta$ either a fermionic or bosonic Matsubara frequency 
depending on the type of the operators $\hat A$ 
and $\hat B$. The contour $\mathcal{C}$ encircles all singularities of the integrand, which
are situated on the $\mathrm{Im}\, z =0$ and $\mathrm{Im}(z+i\eta) =0$ axes 
in a mathematicalley positive sense. 
Performing the trace over the $c$-electrons first, the 
reduced $f$-partition function $Z_f$, the 
$f$-electron one-body Green function $F_\s\equiv G_{f_\s,f^\dagger_\s}$
and generalized 
susceptibility $\chi_{M,M'}\equiv G_{\hat{X}_{MM},\hat{X}_{M'M'}}$ 
can be expressed as
\begin{align}
  \label{eq:app-ZP}
  Z_f&=\sum_M\oint_\mathcal{C} \!\frac{dz}{2\pi i}\:e^{-\beta\,z}\:P_M(z)\quad,
\end{align}
\begin{align}
  \label{eq:app-GP}
  F_{\s}(i\omega_n)&=\frac{1}{Z_f}\oint_\mathcal{C} \!\frac{dz}{2\pi i}\:e^{-\beta\,z}  \Big[
  \\ \notag
  & 
  \phantom{+}
  P_0(z)\:P_{\s}(z+i\omega_n)\:\Lambda_{0,\s}(z,i\omega_n)
  \\\notag
  & 
  +P_{\bar{\s}}(z)\:P_{2}(z+i\omega_n)\:\Lambda_{2,\bar{\s}}(z+i\omega_n,i\omega_n)
  \:\Big]
  \quad,
\end{align}
\begin{align}
  \chi_{M,M'}(i\nu_n)&=
  \notag
    -\frac{1}{Z_f}\oint_\mathcal{C}\!\frac{dz}{2\pi i} 
    \:e^{-\beta z}\:\chi_{M,M'}(z,z+i\nu_n)
    \\ 
    &=-\frac{1}{Z_f}\oint_\mathcal{C}\!\frac{dz}{2\pi i} 
    \:e^{-\beta z}\:\Gamma_{M,M'}(z,z+i\nu_n)\:\cdot
    \notag
    \\
    &  \label{eq:app-chiGammaPPDef}
  \phantom{-\frac{1}{Z_f}\oint_\mathcal{C}}\cdot
  \Pi_{M'}(z,z\!+\!i\nu_n)
  \quad.
\end{align}
Here  $\Pi_M(z,z')=P_M(z)P_M(z')$ and the ionic propagators
\begin{align}
  \label{eq:app-ionicProp}
  P_M(z)&=\bra{M}\ew{\klammerb{[}{]}{z-\hat{H}+\hat{H}_c}^{-1} }_{c}\ket{M}
  \\
  \label{eq:app-ionicSelfenergy}
  &\equiv\frac{1}{z-E_M-\Sigma_M(z)}
  \quad,
\end{align}
which describe the dynamics of an ionic state $\ket{M}$, are introduced.
In equation (\ref{eq:app-ionicSelfenergy})
the ionic propagator is expressed  with the help of the ionic self-energy 
$\Sigma_M(z)$.
$\ew{\dots}_c=\frac{1}{Z_c}Tr_c\big(\:e^{-\beta\hat H_c}
\dots\big)$ indicates that the trace is to be taken over the $c$-states only, and  
$Z_c$ represents the partition function of the isolated $c$-system. 
In equations (\ref{eq:app-GP}) and (\ref{eq:app-chiGammaPPDef}), 
$\Lambda_{M,M'}$ and $\Gamma_{M,M'}$ represent vertex 
functions to be specified later. These equations
are graphically represented in Figure \ref{fig:appNCABase}. 

After re-writing the Hamiltonian in terms of 
the ionic transfer operators $\hat X_{M,M'}$,
the resolvent operator is expanded with respect to $\hat V$,
$  \klammerb{[}{]}{z-\hat{H}}^{-1}=\klammerb{[}{]}{z-\hat{H}_f-\hat{H}_c}^{-1}
\sum_{n=0}^\infty\klammer{[}{]}{\hat{V}\,\klammerb{[}{]}{z-\hat{H}_f-\hat{H}_c}^{-1}}^n
$.
Consequently inserting the identity $\hat{1}=\sum_{c;M}\ket{M}\ket{c}\bra{c}\bra{M}$
and using the representation (\ref{eq:app-ionicProp}) for the ionic propagators,
this perturbation theory can be formulated 
with time-ordered Goldstone diagrams, representing the dynamics 
of the ionic states $\ket{M}$ and their hybridization processes\footnote{
The ionic transfer operators can be avoided by enlarging the Hilbert 
space and introducing  auxiliary slave-bosons, which represent the empty 
state, see Ref~\onlinecite{colemanSlaveBoson84}. 
The resulting standard Feynman perturbation-theory is in a one-to-one
correspondence to the  non-standard time-ordered Goldstone expansion 
described  here.}.

Approximations are then introduced for the self-energies $\Sigma_M$ and
vertex functions $\Lambda_{M,M'}$ and $\Gamma_{M,M'}$. In order to be able to describe 
non-perturbative many-body phenomena like the Kondo effect, certain classes of diagrams
have to be re-summed up to infinite order, resulting in a formulation
in terms of skeleton diagrams, and corresponding coupled integral equations for
the relevant dynamic functions. Deriving these equations consistently through functional derivatives 
from one Luttinger-Ward functional $\Phi$ renders these approximations
thermodynamically consistent.

\subsection{ENCA for generalized dynamic susceptibilities}
\label{sec:encaSus}

Within the ENCA, the ionic self-energies and propagators have to be determined from
the well known set of  coupled integral equations,\cite{pruschkeENCA89,keiterENCA90,sasoFiniteUNCA92,greweCA108}
\begin{align}
  \label{eq:app-sigNCA}
  \Sigma_{0}(z)&=\sum_\s\int\!dx\:\Delta^+(x)\:P_{\s}(z+x)\:\Lambda_{0,\s}(z,x)
  \\\notag
  \Sigma_{\s}(z)&=\int\!dx\:\Delta^-(x)\:P_{0}(z-x)\:\Lambda_{0,\s}(z-x,x)
  \\\notag
  &\phantom{=}+\int\!dx\:\Delta^+(x)\:P_{2}(z+x)\:\Lambda_{2,\s}(z+x,x)
  \\\notag
  \Sigma_{2}(z)&=\sum_\s\int\!dx\:\Delta^-(x)\:P_{\s}(z-x)\:\Lambda_{2,\s}(z,x)
  \quad,
\end{align}
where the vertex functions are given by
\begin{align}
  \label{eq:app-vertexENCA}
  \Lambda_{0,\s}(z,z')&=1+\int\!dx\:\Delta^+(x)\:P_{\bar{\s}}(z+x)\:P_2(z+z'+x)
  \\\notag
  \Lambda_{2,\s}(z,z')&=1+\int\!dx\:\Delta^-(x)\:P_{\bar{\s}}(z-x)\:P_0(z-z'-x)
  \quad,
\end{align}
and $\Delta^\pm(x)=\Delta(x)f(\pm x)$.  The
hybridization function $\Delta(x)$ is constructed from
the $c$-band electrons
\begin{align}
  \label{eq:app-mediumDef}
  \Delta(x)&=\frac{1}{N_0}\sum_{\underline{k}}|V_{\underline{k}}|^2\:\delta(x-\e^c_{\underline{k}})
  \quad,
\end{align}
and  $f(x)=1/(e^{\beta x}+1)$ is the Fermi function with $\beta=1/T$ 
the inverse temperature (note $\hbar=c=k_B=1$).

\begin{figure*}[t]
  \includegraphics[scale=1]{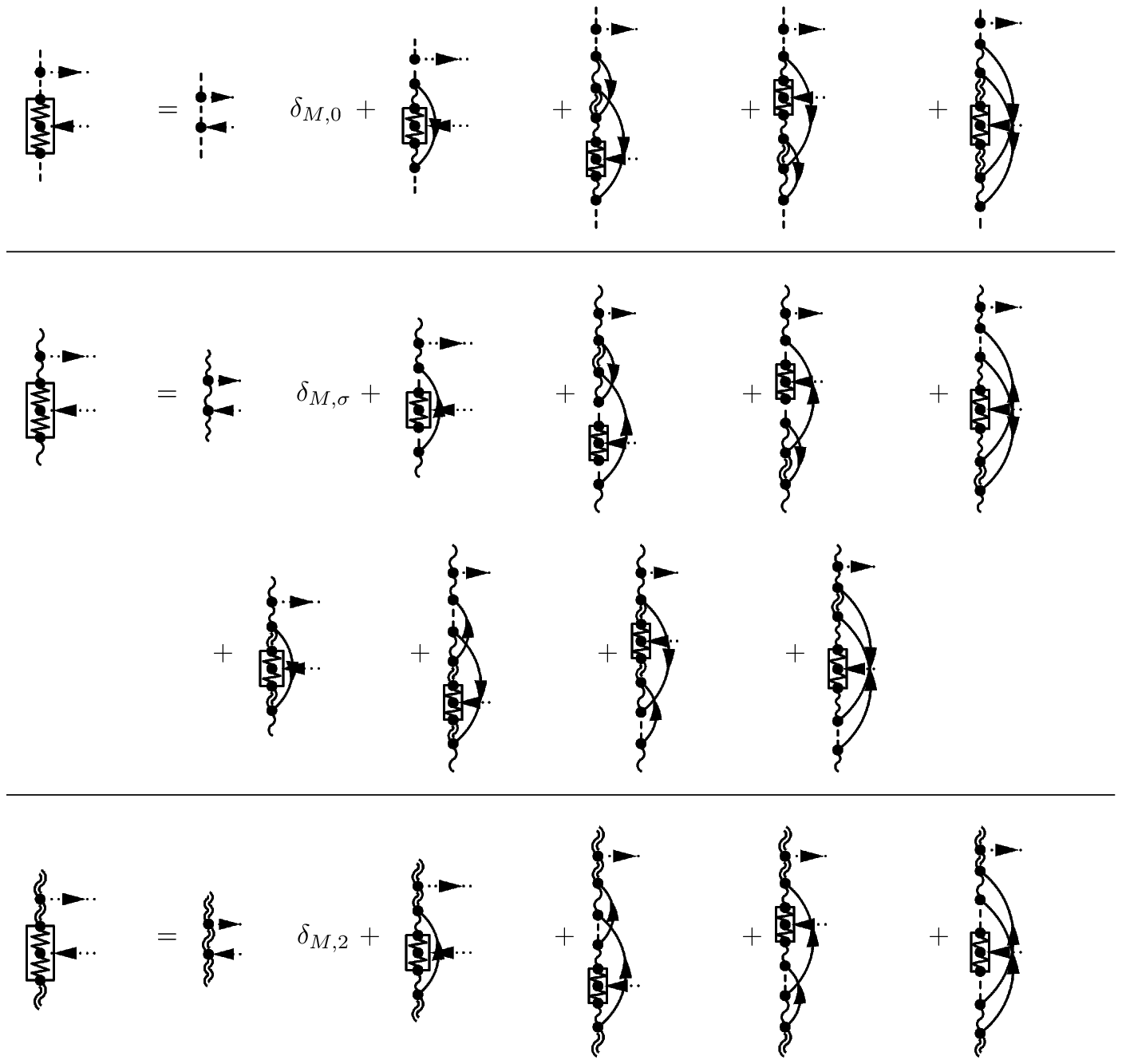}
  \caption{ System of integral equations for $\chi_{M,0}(z,z')$, $\chi_{M,\sigma}(z,z')$, 
    and $\chi_{M,2}(z,z')$ within the ENCA. 
    Zig-zag lines in the vertex part represent the ionic state $M$, 
    which can either be empty, singly occupied or doubly occupied. The arrows entering 
    and leaving each diagram carry the external bosonic frequency $i\nu$. 
 }
  \label{fig:appChiENCAM}
\end{figure*}

For the  generalized susceptibilities $\chi_{M,M'}(z,z')$, an 
additional set of integral equations has to be set up
which is shown graphically 
in Figure \ref{fig:appChiENCAM}.
In principle, there are 16 such functions, but for each quantum number $M$  only
the four functions  $\chi_{M,0}(z,z')$, $\chi_{M,\sigma}(z,z')$, 
and $\chi_{M,2}(z,z')$ are coupled. 
Due to the conserving nature of the ENCA, these equations are closely 
related to the ENCA expressions for the ionic self-energies; only some
additional bosonic lines entering and leaving the site have to be introduced.  

Whereas the lowest order vertex corrections of equation 
(\ref{eq:app-vertexENCA}) together with the 
ionic propagators (\ref{eq:app-ionicProp}) and self-energies 
(\ref{eq:app-ionicSelfenergy}), are sufficient to furnish a conserving
approximation, the vertices
$\Gamma_{M,M'}(z,z')$ for the susceptibilities (\ref{eq:app-chiGammaPPDef})
have to be iterated up to infinite order!
The transcription of these graphs into formulas 
is straight forward but lengthy and will be omitted here.

For very large Coulomb interactions the spectral weight in the doubly-occupied  propagator
is completely moved to energies of the order of $U$. This leads to a negligible 
influence in the region of accessible energies of the order of the bandwidth
due to the large energy denominators,
and all terms involving $P_2$ effectively drop out of the equations.
Therefore, for infinitely large Coulomb repulsion $U\to\infty$,
the ENCA reduces to the usual NCA and all equations presented above 
approach the ones already known from the literature.\cite{Kuramoto:ncaI83,Grewe:siam83}
In this sense,  the ENCA is still referred to as 
non-crossing even though crossing diagrams are included. 
For approximations involving ``real'' crossing diagrams see
Refs.~\onlinecite{anders:PostNCA95,greweCA108,kroha:CTMA97,kroha:NCA_CMTA_rev05}.
 
The functions $\chi_{M,M'}(z,z')$ fulfill the symmetry relations 
\begin{align}
  \label{eq:app-symChi}
  \chi_{M,M'}(z^*,{z'}^*)&={\chi_{M,M'}(z,z')}^*
\end{align}
and
\begin{align}
  \label{eq:app-symChi2}
  \chi_{M,M'}(z,z')&=\chi_{M,M'}(z',z)
  \quad, 
\end{align}
which are revealed by inspection of the perturbation expansion.
The susceptibilities also  obey the sum rules
\begin{align}
  \label{eq:susSumRuleChi}
  \sum_{M'}\chi_{M',M}(i\nu_n)&=
  \ew{\hat{X}_{MM}} \:\delta_{i\nu_n,0}
  \\\label{eq:susSumRuleChiA}
  \sum_{M}\chi_{M',M}(i\nu_n)&=
  \ew{\hat{X}_{M'M'}} \:\delta_{i\nu_n,0}
  \\
  \label{eq:susSumRuleXi}
  \sum_{M,M'}\chi_{M',M}(i\nu_n)&=  \sum_M\ew{\hat{X}_{MM}} \:\delta_{i\nu_n,0}
  =\delta_{i\nu_n,0}
  \quad,
\end{align}
which arise from the completeness relation of the local ionic states,
$1_f=\sum_M\ket{M}\bra{M}$.
These sum rules transform into equivalent statements
for the functions $\chi_{M',M}(z,z')$,  
\begin{align}
  \label{sumRuleS}
  \sum_{M'}\chi_{M',M}(z,z')  
  &=-\frac{P_M(z)-P_M(z')}{z-z'}
  \quad.
\end{align}
It can be analytically checked that the  ENCA 
respects these sum rules. The form of equation 
(\ref{sumRuleS}) explicitly reveals the conserving nature of
this approximation, since the sum of sets of two-particle correlation 
functions is determined by the corresponding one-particle correlation function, i.e.\
the ionic propagator. Insofar the equations (\ref{sumRuleS}) resemble
generalized Ward-Identities.

In order to obtain the physical susceptibility as a function of a real frequency, 
the contour integration of equation~\eqref{eq:app-chiGammaPPDef} has to be performed,
and then the external  bosonic Matsubara frequency can be analytically continued 
to the real axis,  $i\nu_n\to\nu\pm i\delta\equiv \nu^\pm$ ($\delta>0$ infinitesimal).
The result of this procedure reads
\begin{align}
  \label{eq:app-chiFinal}
  \chi_{M,M'}(\nu^{+})&=
  \int_{-\infty}^\infty\!\frac{d\omega}{2\pi i}
  \Bigg[
  Y_{M,M'}(\omega,\omega\!+\!\nu) 
  \\\notag
  &\phantom{M\equiv}
 \phantom{  \int_{-\infty}^\infty\!\frac{d\omega}{2\pi }}
  -Y_{M,M'}(\omega,\omega\!-\!\nu)^*
  \Bigg]
  \quad,
\end{align}
where the symmetry relations~\eqref{eq:app-symChi} were used and
the defect quantities 
\begin{align}
  \label{eq:defectSus}
  &Y_{M,M'}(x,y) = 
  \frac{e^{-\beta x}}{Z_f}
  \klammerB{[}{]}{
    \chi_{M,M'}(x^+,y^+)-
    \chi_{M,M'}(x^-,y^+)
  }
\end{align}
were defined. 
Due to the appearance of the exponential factor,
a direct numerical evaluation of the defect 
quantities given the $\chi_{M,M'}$ is not possible,
and additional sets of integral equations have to
be solved for the $Y_{MM'}$ (cf.\ Ref.~\onlinecite{Kuramoto:ncaIII84} and 
\onlinecite{Kuramoto:ncaIV86}). 

In the following, only spin-symmetric situations are considered, and the
propagators for opposite spins are identified, i.e.\ 
$P_{\s}=P_{\bar{\s}}\equiv P_{\uparrow}$, 
$\chi_{\s,\s}=\chi_{\bar{\s},\bar{\s}}\equiv\chi_{\uparrow,\uparrow}$,
$\chi_{\bar{\s},\s}=\chi_{\s,\bar{\s}}\equiv\chi_{\uparrow,\downarrow}$ \dots.

\subsubsection{\label{sec:app-MagSusNCA} Magnetic susceptibility}

The relevant quantity for the magnetic susceptibility is the auto-correlation function 
of the $z$-component of the spin operator 
$\hat S^z \sim \hat n_\uparrow-\hat n_\downarrow= \hat X_{\uparrow,\uparrow}
-\hat X_{\downarrow,\downarrow}$. This translates into the linear combination
\begin{align}
  \label{eq:app-ncaMagSusOp}
  \chi_{mag}(z,z')&=  \chi_{\uparrow,\uparrow}(z,z')-\chi_{\uparrow,\downarrow}(z,z')
 \quad,
\end{align}
which needs to be determined. Setting up the equations for this linear combination
using the general equations depicted in Figure \ref{fig:appChiENCAM}
leads to the cancellation of all spin-symmetric terms.\cite{schmittPhD08}
The function $\chi_{mag}(z,z')$
decouples from the $\chi_{0,M}(z,z')$ and $\chi_{2,M}(z,z')$,
and only one integral equation results,
\begin{align}
  \label{eq:app-nca-chiMag}
  &\chi_{mag}(z,z')=
  \Pi_{\uparrow}(z,z')\Big\{ 
  \\\notag  &\phantom{MM} 
  1-\int\!dx dy\Delta^+(x)\Delta^-(y)\Big[
  P_{2}(z'\!+\!x) \:P_0(z\!-\!y) 
  \\\notag 
  &\phantom{MM}
  + P_{0}(z'\!-\!y) \:P_{2}(z\!+\!x)
  \Big]
  \:\chi_{mag}(z\!+\!x\!-\!y,z'\!+\!x\!-\!y)
  \Big\}
  \quad.
\end{align}
The derivation of the corresponding defect equation for $Y_{mag}$ 
along the lines of the definition (\ref{eq:defectSus}) is straight forward,  
but will be omitted here for brevity.

\subsubsection{ \label{sec:app-ChargeSusNCA} Charge susceptibility}

For the charge susceptibility the relevant quantity is the auto-correlation function 
of the charge operator  $\hat n=\hat n_\uparrow+\hat n_\downarrow= \hat X_{\uparrow,\uparrow}
+\hat X_{\downarrow,\downarrow}+2\hat X_{2,2}$ leading to the dynamic function 
\begin{align} 
  \label{eq:app-ncaCharegeCmp}
  &  \chi_{charge}(z,z')=  \chi_{\uparrow,\uparrow}(z,z')+\chi_{\uparrow,\downarrow}(z,z')
  \\\notag&
  \phantom{MMM}
  +2\klammerB{(}{)}{  
    \chi_{2,2}(z,z')+\chi_{2,\uparrow}(z,z')+\chi_{\uparrow,2}(z,z')
  }
 \\\notag 
   &\phantom{M}=\chi_{2,2}(z,z')-\chi_{0,2}(z,z')
  +\chi_{0,0}(z,z')-\chi_{2,0}(z,z')
  \\\notag
  &  \phantom{MM} +S_{0,\uparrow,2}(z,z')
  \quad.
\end{align}
In the second equality sum rules like (\ref{sumRuleS}) were used.
The function $S_{0,\uparrow,2}(z,z')$ is not relevant for dynamic
susceptibilities,
since it only contributes in the case of a vanishing external
frequency $i\nu_n=0$. 

The hole susceptibility, i.e.\ the auto-correlation function  
of the hole operator $\hat h=2-\hat n= 2\hat X_{0,0}+\hat X_{\uparrow,\uparrow}
+\hat X_{\downarrow,\downarrow}$, is given by the same expression, only with
a different contribution from the sum rules.  Both of these contributions do not
change the dynamic susceptibility since they produce only
the necessary constant shift between the static ($i\nu_n=0$) charge and
hole susceptibility after the contour integration,
\begin{align}
  \label{eq:app-nca-ChargeSusFreq}
  \chi_{hole}(i\nu_n)&= \chi_{charge}(i\nu_n) +4(1-\ew{\hat n})\delta_{i\nu_n,0}
  \quad .
\end{align}

The dynamical quantities of interest are best obtained by setting up the integral 
equations for the 
linear combinations $c_{0,0}=\chi_{0,0}-\chi_{2,0}$, 
$c_{0,\uparrow}=\chi_{0,\uparrow}-\chi_{2,\uparrow}$
and $c_{0,2}=\chi_{0,2}-\chi_{2,2}$, which read
\begin{align}
&  c_{0,0}(z,z')= \Pi_{0}(z,z')\Bigg\{
  1  +2 \int\!dx\Delta^+(x)\:\Big[\Lambda_{0,\uparrow}(z',x)
\notag \\ &   \label{eq:app-charge00} 
\phantom{MMMM}+\Lambda_{0,\uparrow}(z,x)-1\Big]
  \:c_{0,\uparrow}(z\!+\!x,z'\!+\!x)
  \\\notag
&  +2\int\!dx\,dy\Delta^+(x)\Delta^+(y)
  \:P_{\uparrow}(z'\!+\!x) \: P_{\uparrow}(z\!+\!y)
  \times   \\\notag
  &\phantom{MMMM}
  \times c_{0,2}(z\!+\!x\!+\!y,z'\!+\!x\!+\!y)
  \Bigg\}
\end{align}
\begin{align}
  \label{eq:app-charge10}
  & c_{0,\uparrow}(z,z')=\Pi_{\uparrow}(z,z')\Bigg\{ 
  \\\notag
  &
  \int\!dx\Delta^-(x)\Big[ \Lambda_{0,\uparrow}(z'\!-\!x,x)
  \\\notag &
  \phantom{MMMM}
  +\Lambda_{0,\uparrow}(z\!-\!x,x)
  -1\Big]\:c_{0,0}(z\!-\!x,z'\!-\!x) 
\\  \notag 
  \shoveright{
    &   + \int\!\!dx\Delta^+(x)\big[\Lambda_{2,\uparrow}(z'\!+\!x,x)
   \\\notag
   &  \phantom{MMMM}
    +\Lambda_{2,\uparrow}(z\!+\!x,x)
   -1\big]
   \:c_{0,2}(z\!+\!x,z'\!+\!x)
  }
  \\
  \notag
  &  +\int\!dx dy\Delta^+(x)\Delta^-(y)\Big[
    P_{2}(z'\!+\!x) \:P_0(z\!-\!y) 
    \\
    \notag
    &\phantom{MMM}
    + P_{0}(z'\!-\!y) \:P_{2}(z\!+\!x)
  \Big]
  \:c_{0,\uparrow}(z\!+\!x\!-\!y,z'\!+\!x\!-\!y)
  \Bigg\}
\end{align}

\begin{align}
  &c_{0,2}(z,z')= \Pi_2(z,z')\Bigg\{
  -1
  + 2 \int\!\!dx\Delta^-(x)\Big[ \Lambda_{2,\uparrow}(z',x)
  \notag \\ &  \label{eq:app-charge20}
  \phantom{MMMM}+ \Lambda_{2,\uparrow}(z,x)-1\Big]
  \: c_{0,\uparrow}(z\!-\!x,z'\!-\!x)
  \\\notag 
  &  +2\int\!dx dy\Delta^-(x)\Delta^-(y) 
  \:P_{\uparrow}(z'\!-\!x) \:P_{\uparrow}(z\!-\!y) \:
  \times
  \\\notag
  &\phantom{MMMMMM}
  \times c_{0,0}(z\!-\!x\!-\!y,z'\!-\!x\!-\!y)
  \Bigg\}
    \quad.
\end{align}
All three linear combinations are now coupled which makes the 
numerical solution of the full system inevitable.  

The corresponding defect equations are again determined in a
straight-forward way,
but are omitted here due to their length. 

\subsection{Numerical evaluation}
The ionic propagators and defect propagators are strongly peaked 
at the ionic threshold, and they even develop a singularity at zero 
temperature.\cite{greweCA108,Kuramoto:ncaIII84,muellerhartmann:NCAgroundstate84}
The position of this ionic threshold is not know exactly a priori, so we 
use self-generating adaptive frequency meshes to represent all functions
numerically. 

After obtaining the ionic propagators from the set of integral 
equations Eq.~(\ref{eq:app-sigNCA}),
Eqs.~(\ref{eq:app-nca-chiMag}) and (\ref{eq:app-charge00}) -- (\ref{eq:app-charge20})
can be solved to yield the two-particle quantities $\chi_{mag}(\omega^\pm,\omega\pm\nu^+)$
and $c_{M,M'}(\omega^\pm,\omega\pm\nu^+)$. With these at hand, the corresponding defect equations
can be solved and physical susceptibilities along the lines of Eq.~(\ref{eq:app-chiFinal})
can be extracted.

All integral equations are solved via Krylov subspace methods,\cite{saadIterativeMethodsSparseLineSys_Book}
where --- starting from a sophisticated guess for the unknown functions ---
the equations are iterated until convergence is reached. 
In order to accelerate convergence and stabilize the iteration procedure
Pulay-mixing\cite{pulayIterativeConvergence80} between different iterations is implemented (see, for example, 
Ref.~\onlinecite{johnsonConvergenceCalc88} or \onlinecite{eyertConvergence96}).

Unfortunately, the system for the  defect quantities becomes singular for 
vanishing external frequency $\nu\to 0$. This is seen at the sum rule
(\ref{sumRuleS}), which translates into  $\sum_{M'}Y_{M',M}(\omega,\omega+\nu)=-2\pi i \xi_M(\omega)/(\nu+i\delta)$
implying that the $Y_{M,M'}$ become of the order of $\frac{1}{\nu}$ for small $\nu$,
while all terms in the integral equations stay at the order one.
This becomes of some importance when extracting static
susceptibilities from calculations of  the dynamic susceptibilities, where  a very small but finite frequency
is used. The resulting convergence problems are probably connected to the ones 
already mentioned in Ref.~\onlinecite{brunner:ChargeSusSIAM97}.
More details on the numerics can be found in Ref.~\onlinecite{schmittPhD08}.

\section{\label{sec:app-ncaPhysSIAM} Results }

\begin{figure}
  \includegraphics[scale=0.65]{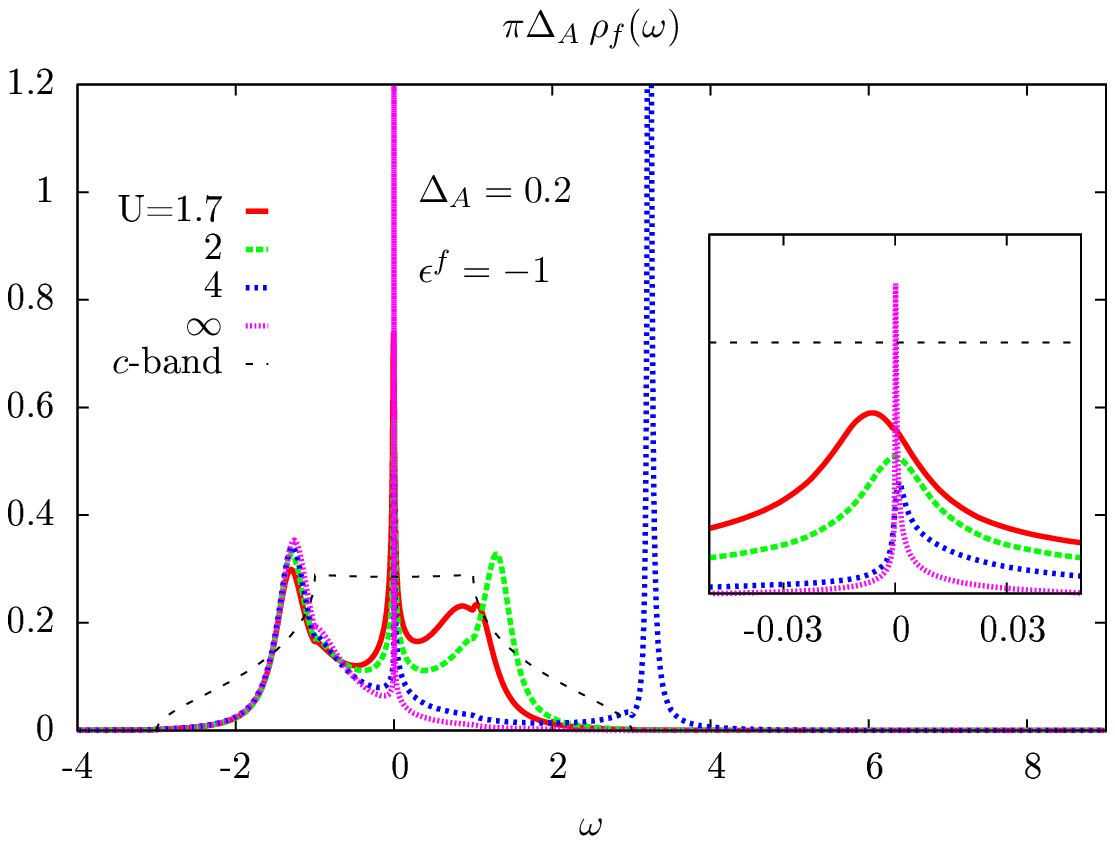}\\ 
  \includegraphics[scale=0.65]{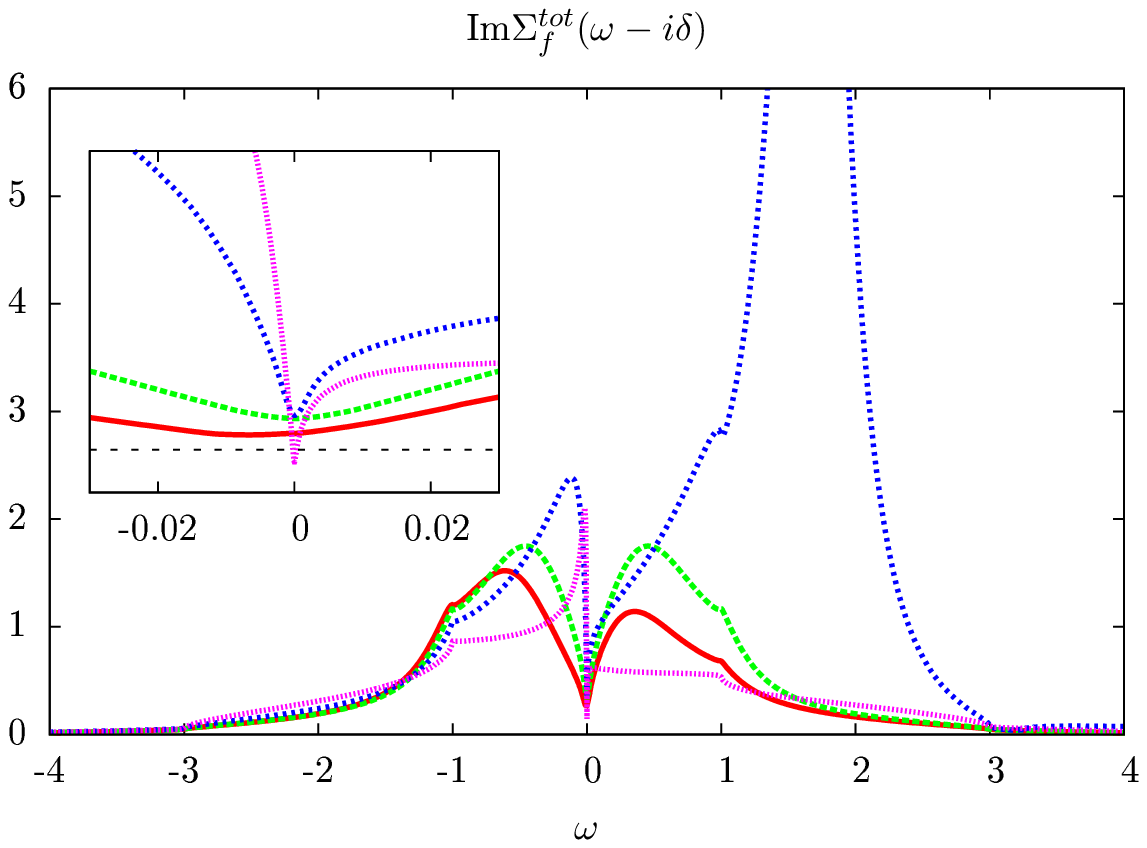}\\
  \caption{ \label{fig:app-nca_gf_asym_siam}
    Rescaled $f$-electron spectra (upper graph)
    and total self-energy (lower graph) as a function of energy for various
    values of the Coulomb repulsion $U$. 
    The temperature was fixed at half the Kondo temperature $T\approx T_K/2$ of each case,
    the ionic level was chosen $\e^f=-1$ and the Anderson width was $\Delta_A=0.2$.
    The conduction band DOS was that of a three-dimensional tight binding lattice
    with half bandwidth $W=3$ (``$c$-band'' in the upper graph).
    The insets show the low energy interval around the Fermi level, where the
    energy is measured in units of the  corresponding Kondo temperature and the 
    horizontal lines indicate the $T=0$ limiting values. 
  }
\end{figure}

As it is well known, the SIAM exhibits the Kondo effect for sufficiently
large Coulomb repulsion $U$ and the single particle level below the 
Fermi energy, $\e^f<0$.
A very prominent manifestation of that effect is the emergence of the Abrikosov-Suhl resonance (ASR)
in the one-particle density of states (DOS) at the Fermi level 
for temperatures of the order of the Kondo temperature $T_K$. 
Within the ENCA an order of magnitude 
estimation for $T_K$ is given by~\cite{pruschkeENCA89}
\begin{align}
  \label{eq:app-nca_TKENCA}
  T_K&=\frac{\mathrm{min}(W,U)}{2\pi} \sqrt{J}\:e^{-\pi/J}
 \:,\:\: J=-\frac{2U\Delta_A}{\e^f(\e^f+U)}
\end{align}
where $J$ is  the effective antiferromagnetic Schrieffer-Wolff exchange coupling,
the Anderson width  $\Delta_A=\pi\rho_c(0) V^2$ and $W$ the half bandwidth
of the conduction electron DOS $\rho_c(\omega)$. The hybridization matrix element
was assumed to be local, i.e.\ momentum independent $V_{\underline{k}}\to V$.

As already mentioned, the pathology of the NCA
manifests itself in the overestimation of the height of
the ASR and a violation of Fermi liquid properties for too low 
temperatures as well as in in situations with large valence 
fluctuations.\cite{Grewe:siam83,Kuramoto:ncaIII84,kuramoto:AnalyticsNCA85,bickers:nca87a}
This pathological behavior is strongest for the case
of a spin-only  degeneracy ($N=2$), which  is considered in this work.

Figure \ref{fig:app-nca_gf_asym_siam} shows the one-particle 
$f$-electron DOS  $\rho_f(\omega)=\frac{1}{\pi}\mathrm{Im}\,F_\s(\omega-i\delta)$
and the imaginary part of the total self-energy
Im$\Sigma^{tot}_f(\omega-i\delta)=-\mathrm{Im}[1/F_\s(\omega-i\delta)]$
calculated within the ENCA and the NCA ($U=\infty$).
As can be seen, the ENCA does not 
overestimate the height of the Kondo peak
or violate the Fermi liquid property
of the total self-energy Im$\Sigma^{tot}_f(0-i\delta)\geq \Delta_A$
down to temperatures of half the
Kondo temperature. In contrast, the $U=\infty$ NCA curves in the graphs
do violate  these limits for the same parameter values.

The fact that the ENCA performs better than the NCA 
in comparable situations
is a consequence of a better balance between different kinds of perturbational
processes.  The importance of such a balance 
has been pointed out repeatedly.\cite{greweCA108,keiterGeneralizedMagImp85,keiterIGLSIAM90}

Even though the performance of the ENCA is considerably enhanced over the NCA, it 
eventually overestimates the height of the ASR for even lower temperatures
and still misses to produce the correct $T=0$ Fermi liquid relations. 
Further improvements can be archived via the incorporation of higher-order
diagrams.\cite{greweCA108}

\subsection{\label{sec:Bench} Benchmarking the ENCA}
\subsubsection{\label{sec:app-ncaPhysSIAMSymm} Static susceptibilities in the symmetric case}
\begin{figure}
  {  \scriptsize
    \includegraphics[scale=0.65]{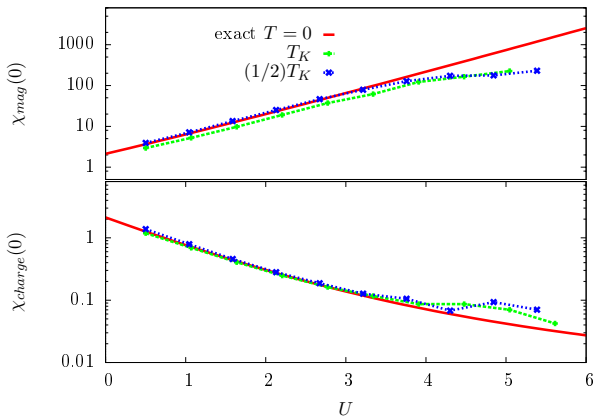}
  }
  \caption{Static magnetic (upper) and charge (lower) susceptibility for the 
    symmetric case $\e^f=-U/2$ as a function of $U$ for  finite 
    temperatures $T_K$ and $T_K/2$ in comparison to  the exact $T=0$ results 
    of equations~\eqref{eq:app-ncaExactSusMag} and~\eqref{eq:app-ncaExactSusCha}. 
    The calculations were done for a constant $c$-electron DOS with half bandwidth $W=10$
    and an Anderson width $\Delta_A=0.3$.
  }
  \label{fig:app-nca_staticChiSymExact}
\end{figure}

In order to obtain a better understanding of possible shortcomings of the ENCA it is worthwhile to consider charge and spin excitations separately
and benchmark them against some exactly known results. 
The thermodynamics of the SIAM can be obtained exactly from the Bethe ansatz 
method.\cite{kawakamiGroundStateSIAM81,wiegmannExactSIAMI83,
  tsvelickExactSIAMII83,kawakamiSIAMExactHeat83,okijiSIAMExactSus83}
At zero temperature and for the symmetric case ($\e^f=-\frac{U}{2}$) with a flat conduction band of infinite
bandwidth ($W\to\infty$) the static 
susceptibilities can even be obtained in closed form,\cite{okijiSIAMExactSus83,horvaticExactSIAMPerturb85}
\begin{align}
  \label{eq:app-ncaExactSusMag}
  \chi_{mag}^{exact}(T=0)&=\frac{(g\mu_B)^2}{4\,T_L}
  \Bigg(
    1+\frac{1}{\sqrt{\pi}}\int_0^{\frac{\pi\Delta_A}{2U}}\!\!dx\:\frac{e^{\big\{x-\frac{\pi^2}{16x}\big\}}}{\sqrt{x}}
  \Bigg)
\end{align}
with 
\begin{align}
  \label{eq:app-ncaExactSusTL}
  T_L&=U\sqrt{\frac{\Delta_A}{2U}}\:\exp\left\{-\frac{\pi U}{8\Delta_A}+\frac{\pi\Delta_A}{2U} \right\}
\end{align}
and 
\begin{align}
  \label{eq:app-ncaExactSusCha}
  \chi_{charge}^{exact}(T=0)&=\frac{1}{\pi}\sqrt{\frac{2}{U\Delta_A}}\int_{-\infty}^\infty\!\!dx\:
  \frac{
    e^{-\frac{\pi\Delta_A}{2U}\,x^2}
  }{1+\left(\frac{U}{2\Delta_A}+x\right)^2}
  \quad.
\end{align}
Apart from the small coupling correction $\sim\frac{\Delta_A}{U}$ in the exponent,
$T_L$ exactly coincides with the Kondo temperature of equation~(\ref{eq:app-nca_TKENCA}). 

In the graphs of Figure \ref{fig:app-nca_staticChiSymExact} the
exact zero temperature magnetic (upper) and charge (lower) susceptibilities 
of equations~\eqref{eq:app-ncaExactSusMag} and 
\eqref{eq:app-ncaExactSusCha} (solid red lines) are compared  to
the ENCA susceptibilities (blue dots)
for $\e^f=-U/2$ as a function of $U$.  For the ENCA curves
two characteristic  temperatures $T_K$ and $T_K/2$ are chosen.

The characteristic exponential $U$ dependence of the magnetic susceptibility 
is essentially the same as for the exact  Bethe ansatz result, but the absolute 
height is somewhat different. At $T=T_K$ the magnetic susceptibility is
roughly $70\%$  of its zero temperature saturation value. However,
at $T=T_K/2$
the susceptibility has almost saturated and the agreement
with the $T=0$-value is quite good.

The deviations for $U>4$  are due to the method used to extract the 
static susceptibilities: The static limit is obtained by evaluating 
the dynamic susceptibility at a small but finite  external
frequency, in our case $\nu=10^{-5}$. 
Since the ENCA 
represents a conserving approximation, the results obtained with this
method agree with 
the static susceptibility obtained from a derivative of a thermodynamic
potential, or solving separate
equations as in \citet{otsukiNCAThermodyn06}.
This is valid as long as the
minimal frequency is negligible compared to the lowest energy scale in the problem. 

However, the Kondo temperature 
for $U=4$ is only of the order of $10^{-4}$ and decreases
for larger $U$.
The susceptibility calculated at $\nu=10^{-5}$  then does no longer represent the
static limit anymore, and the  decrease seen in 
Figure \ref{fig:app-nca_staticChiSymExact} is  produced, 
which is therefore not indicating a shortcoming of the ENCA method. 
The deviation can be cured by 
choosing a smaller value for the external frequency.

The charge susceptibility (lower graph in Figure~\ref{fig:app-nca_staticChiSymExact}) 
shows no significant temperature  dependence for $T=T_K$ and $T_K/2$
and lies right on top of the exact $T=0$ result. The deviations for $U\geq 4$ 
are explained in the same way as for the magnetic susceptibility described above.

\begin{figure}
  {  \scriptsize
    \includegraphics[scale=0.65]{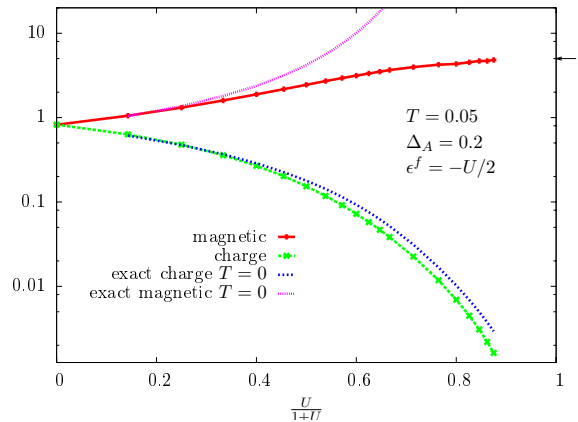}
  }
  \caption{Magnetic and charge susceptibility  for the symmetric 
    impurity ($\e^f=-U/2$) as a function of the
    Coulomb interaction $U$ at a fixed temperature $T=0.05$
    and for an Anderson width $\Delta_A=0.2$. 
    Notice, the $x$-axis is scaled as $U/(1+U)$, so that $U=\infty$ corresponds to $U/(1+U)=1$.
    The arrow on the right of the figure corresponds to the value of a Curie 
    susceptibility  $\chi_{mag}/(g\mu_B)^2=1/(4T)=5$ of a free spin at temperature $T=0.05$. 
  }
  \label{fig:app-nca_staticSusU}
\end{figure}

These reasons for the deviation at larger $U$ can be further confirmed by
calculating the static susceptibilities for a fixed finite temperature $T=0.05$ as 
a function of the Coulomb interaction, which are shown in Figure~\ref{fig:app-nca_staticSusU}.
The figure compares the static magnetic (red dots) and charge (blue dots) susceptibilities 
for the symmetric situation with the exact $T=0$ results.

The charge susceptibility  decreases monotonically with increasing Coulomb 
repulsion as expected. It follows the exact zero temperature susceptibility very accurately,
which again confirms its weak temperature dependence in symmetric situations at low 
temperatures. 

The magnetic susceptibility (red dots) does agree with the exact $T=0$ solution for 
low Coulomb repulsions but deviates for $U/(1+U)\gtrsim 1/3$ ($U\gtrsim1/2$). 
This is a finite temperature effect, since  for 
values of $U\gtrsim 1/2$ the Kondo temperature is smaller 
than the chosen finite temperature of $T=0.05$. Consequentlya, 
for $U\gtrsim 1/2$
we are not in the low temperature regime, and the susceptibility  is  
not well described by its $T=0$ value. The magnetic susceptibility does not grow exponentially with
$U$ as for $T=0$, but instead saturates for $U\to\infty$ at the asymptotic value of 
the Curie susceptibility of a free spin,  $\chi_{mag}=1/(4T)=5$, which is indicated by an 
arrow on the  right border of the graph.

\subsubsection{\label{sec:app-ncaPhysSIAMAsymm} Static susceptibilities  in the asymmetric case}

\begin{figure}
  { 
    \scriptsize
    \includegraphics[scale=0.65]{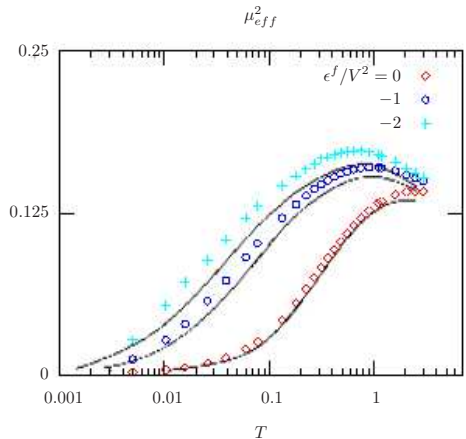}
  }
  \caption{Temperature dependent squared effective magnetic moment within the ENCA (colored
    dots) for a fixed Coulomb repulsion $U=4V^2$ and three different $f$-level 
    positions $\e^f=0,-V^2,-2 V^2$.
    The solid grey curves are the exact Bethe ansatz results for the same parameter values
    (after \citet{okijiThermodynSIAM83}). 
    The calculations were done for a constant $c$-electron DOS with half bandwidth $W=10$
    and $\Delta_A=0.167$.
  }
  \label{fig:app-nca_staticChiMomExact}
\end{figure}

As a first test for the ENCA in the asymmetric situation Figure \ref{fig:app-nca_staticChiMomExact}
compares the square of the effective screened local magnetic moment of the impurity,
\begin{align}
  \label{eq:nca-ScreenedMom}
  \mu_{eff}^2 &
  =\frac{1}{(g\mu_B)^2}T\chi_{mag}(\nu=0)\quad,
\end{align}
to the exact Bethe ansatz result  for three different ionic
level positions as a function of temperature. The solid  grey lines
are the exact Bethe ansatz solution, which is 
taken from Ref.~\onlinecite{okijiThermodynSIAM83}, while
the colored points are ENCA calculations for the same parameter values.
The half bandwidth was taken to be $W=10$, which should be large  enough
to be comparable to the Bethe ansatz solution where  $W=\infty$.

The  ENCA slightly 
overestimates the squared effective moment but all characteristic features are essentially the 
same as for the Bethe ansatz.  Especially the shape and the relative height of
the  curves is in remarkable agreement: All three ENCA curves can be  
brought to lie right on top of the exact Bethe ansatz results when
they are  rescaled with one single factor. This indicates that the ENCA 
produces a slightly modified Kondo scale, but otherwise describes the static
magnetic properties almost exactly.  
This is  especially remarkable for the intermediate valence situation
with $\e^f=0$, where the empty and 
singly occupied ionic configurations are almost degenerate. In such situations
stronger pathologies occur in the one-particle DOS and 
the NCA-type of approximations would be expected to yield results of
lower quality. However, 
as it can be seen, the magnetic excitations are still described very accurately. 

Calculations of the effective squared local moments 
for a wide range of Coulomb interactions and hybridization strength 
(not shown) resemble the exact solutions known from the 
literature~\cite{krishnamurtyNRGSIAMI80,krishnamurtyNRGSIAMII80,okijiSIAMExactSus83}
and the characteristics of the different asymptotic regimes
are well reproduced by the ENCA.

\begin{figure}
  {  \scriptsize
    \includegraphics[scale=0.65]{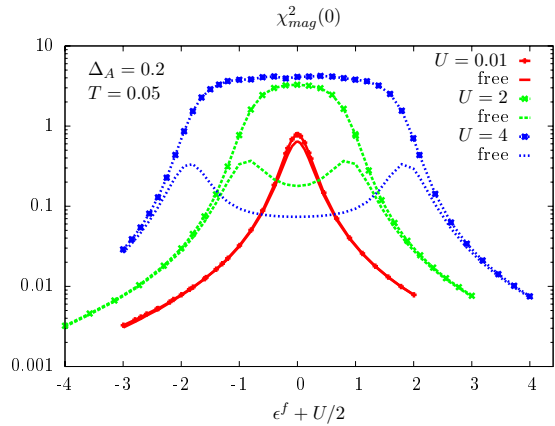}
  }
  \caption{Static magnetic susceptibility
    for a fixed $T=0.05$ and $\Delta_A=0.2$ as functions of the 
    ionic level position $\e^f$ relative
    to the half-filling value $-U/2$ for various values of $U$.
    The conduction band was chosen to be constant with a half bandwidth of $W=10$.
    Curves without dots (``free'')
    are calculated without two-particle interactions, i.e.\ with  the particle-hole propagator
    of equation (\ref{eq:PHprop}).
  }
  \label{fig:app-nca_staticChiSymMag}
\end{figure}

In Figure \ref{fig:app-nca_staticChiSymMag} the magnetic susceptibility is systematically 
examined as a function of the ionic 
level position $\e^f$ and for fixed  $T$ and $U$. 
Also shown are the susceptibilities without explicit two-particle 
interactions  (lines without dots
labeled as ``free''),
i.e.\ the  local particle-hole propagator $P^f(0)$
calculated via
\begin{align}
  \label{eq:PHprop}
  P^f(\nu)&=\int_{-\infty}^\infty\hspace*{-2mm}d\omega\: f(\omega)\rho_f(\omega) 
  \\\notag 
  &\phantom{\int_{-\infty}^\infty}
 \times  \big[F_\uparrow(\omega+\nu+i\delta)
  +F_\uparrow(\omega-\nu-i\delta)\big]
\quad,
\end{align} 
where $F_\uparrow$ represents the one-particle $f$-Green function
(cf.\ equation (\ref{eq:app-GP})), $\rho_f(\omega)$ the corresponding spectrum and 
$f(\omega)$ the Fermi function.

All curves are symmetric around
$\e^f+\frac{1}{2} U=0$ which just reflects the particle-hole 
symmetry of the model.
The particle-hole propagators shows  the expected maxima 
approximately situated  at 
the positions of the Hubbard peaks in the one-particle DOS 
$\e^f+\frac{1}{2}U\approx \pm\frac{1}{2}U$.

For very small $U=0.01$, the susceptibility calculated with the 
ENCA is indistinguishable from the particle-hole propagator, as a
consequence of the near lack of two-particle correlations. 

For larger Coulomb interactions, the ENCA susceptibility shows only one broad maximum around
$\e^f+\frac{1}{2} U=0$ (half filling), which grows in height
and width with increasing $U$. The enhancement of the susceptibility 
is due to the increasing local magnetic moment with larger $U$.
The plateau which develops around zero, is due to
the stability of the local moment
as long as the singly occupied ionic configuration is stable, i.e.\ the
lower Hubbard peak being below 
and the upper above the Fermi level, and the temperature is not too low compared to $T_K$.
But as soon as one of the Hubbard peaks extends over the Fermi level, i.e.\
$\e^f+\frac{1}{2}\Delta_A<0$ or $\e^f+U-\frac{1}{2}\Delta_A>0$, the 
moment is destabilized.
For both Hubbard peaks below (above)  the Fermi level,
the impurity is predominantly doubly occupied (empty) and 
the magnetic susceptibility drops drastically.  
The curve then rapidly approaches the particle-hole propagator,
indicating that
explicit two-particle interactions are unimportant.

The reproduction of the correct results for the effectively 
non-interacting limit at $U=0.01$ as well as the empty- or fully 
occupied regimes is quite remarkable.

\begin{figure}
  {
    \includegraphics[scale=0.65]{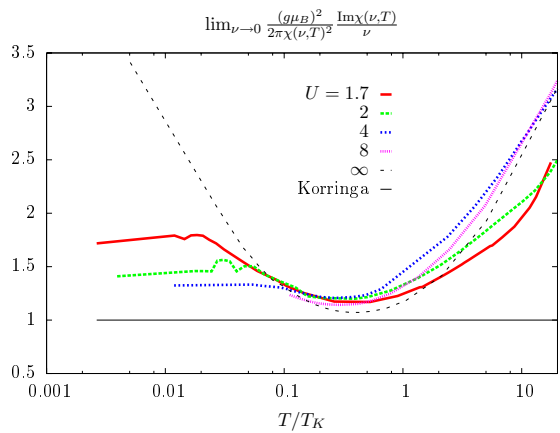}
  }
  \caption{$\lim_{\nu\to0}\frac{(g\mu_B)^2}{2\pi\chi_{mag}(\nu,T)^2}\frac{\mathrm{Im}\chi_{mag}(\nu,T)}{\nu}$
    evaluated as a function of temperature and various Coulomb repulsions within the ENCA. For
    comparison  the $U=\infty$-NCA curve is shown as well.
    A 3d-SC DOS was used for the conduction electrons and $\Delta_A=0.3$.
  }
  \label{fig:app-nca_korringa}
\end{figure}
As it was already mentioned earlier, the NCA does violate Fermi liquid properties
for very low temperatures.
Another indication, in addition to the imbalance of the imaginary part of the  total self energy  at the Fermi 
level (-Im$\Sigma(0+i\delta)<\Delta_A$),
stems from the zero frequency limit of the imaginary part of the dynamic 
susceptibility.  For the Fermi liquid at $T=0$ it has to obey the so called 
Korringa-Shiba relation\cite{shibaKorringaRelationSIAM75}
\begin{align}
  \label{eq:siamKorringa}
  &\lim_{\nu\to0}\frac{(g\mu_B)^2}{2\pi\chi_{mag}(\nu)^2}\frac{\mathrm{Im}\chi_{mag}(\nu)}{\nu}=1
  \quad.
\end{align}
The function on the left hand side of this relation 
is shown in Figure~\ref{fig:app-nca_korringa} 
as a function of temperature for various values of $U$.
The explicit form~(\ref{eq:siamKorringa}) holds for a flat infinitely-wide 
conduction band DOS. For a different DOS the numerical pre-factors might 
change slightly, but the left-hand side is still  expected to be
of the order of unity due to universality of the SIAM at low energies.   

For the NCA, the quantity $\lim_{\nu\to 0}$Im$\chi_{mag}(\nu)/\nu$ is known to diverge 
at $T=0$,\cite{muellerhartmann:NCAgroundstate84} which is reproduced
by the $U=\infty$ curve in the figure.
However, the  ENCA (finite $U$ values)
performs considerably better than the NCA.
The curves still 
slightly increase  for temperatures $T\lessapprox T_K/2$,
but they eventually saturate at a finite value and do not diverge.
This represents a considerable improvement of the qualitative behavior
of the ENCA over the NCA. 


\begin{figure}
  {  \scriptsize
    \includegraphics[scale=0.65]{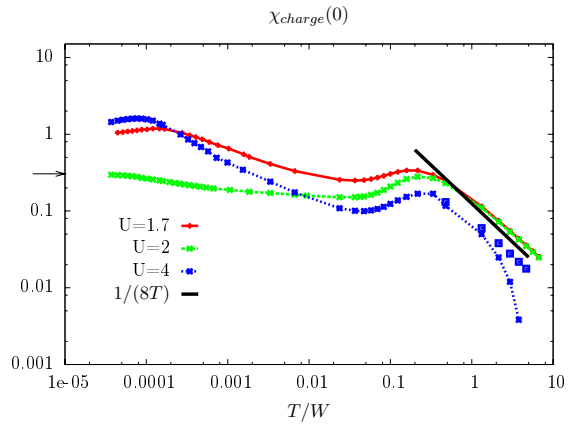}
   } 
  \caption{  \label{fig:app-nca_staticCharge}
    Temperature dependent static charge susceptibility  $\chi_{charge}(0)$ 
    for various Coulomb repulsions $U$ ($\Delta_A=0.2$, $\e^f=-1$).
    The arrow at the left border shows the exact $T=0$
    limit for the symmetric case $U=2$. 
    The high temperature asymptotic form
    of a non-interacting impurity  ($\chi=1/(8T)$) is shown as well (thick line).
    The colored dots, which are not connected by a line, denote the asymptotic 
    susceptibilities calculated without any two-particle interactions, i.e.\ from 
    $P^f(\nu=0)$, see equation (\ref{eq:PHprop}). 
    The conduction band was chosen to be that of a 3d-SC lattice with a half
    bandwidth of $W=3$.
  }
\end{figure}

The temperature dependent static charge susceptibility 
is shown in the 
Figure~\ref{fig:app-nca_staticCharge} 
for various values of $U$ as a function of temperature. 
For high temperatures and $U<4$ 
the static susceptibility behaves effectively non-interacting with 
$\chi_{charge}=1/(8T)$ as expected. 

For $U=4$   the susceptibility still 
has the characteristic  $1/T$ dependence for high $T$, but 
with a  prefactor more closely to $1/16$.
This can be understood since in this situation the upper Hubbard peak 
(incorporating roughly half the spectral weight) is energetically just above the upper  
band edge of the  $c$-band and therfore the accessible spectral weight for
two-particle excitations is approximately halved. 

The rapid drop of the susceptibility for $U=4$
at temperatures $T\gtrsim W$
is attributed to  inaccuracies in the numerics for solving the integral equations.
However, in this effectively non-interacting regime the susceptibility can 
be calculated without explicit two-particle
interactions via the particle-hole propagator
of equation (\ref{eq:PHprop}). The results thus obtained 
are shown in the graph as colored dots without joining lines
and  are seen to be nicely proportional to $1/T$ for high temperatures.

Therefore, the ENCA nicely reproduces the high temperature 
asymptotics of the SIAM. 

At temperatures around $T/W\approx \frac{1}{3}$
all susceptibilities show a pronounced
maximum, which stems from
thermally excited charge fluctuations 
between the empty and singly occupied 
ionic levels with excitation 
energy $\approx|\e^f|/W=1/3$. 
For $U=2$, fluctuations between the singly and doubly occupied 
states have the same excitation energy $|\e^f|/W=(\e^f+U)/W=1/3$ 
and therefore contribute equally. 
For $U=1.7$ the energy of  fluctuations  involving the doubly occupied
state is somewhat smaller ($\sim \frac{0.7}{3} \approx 0.23$)
and the peak is therefore broadened to lower energies. 
For $U=4$ the doubly occupied 
state is inaccessible for thermal fluctuations;
so only the  empty and singly occupied levels 
contribute leading to a reduction of the susceptibility maximum
by approximately a factor of two.

At lower temperatures ($T/W\leq 0.05$ in the figures)
the charge susceptibilities exhibit a slow increase 
followed by a saturation at the zero temperature values. 
The increase in the charge susceptibility occurs in a temperature
range, where the Kondo singlet and
the local Fermi liquid formation take place, which manifests itself 
in the growing many-body resonance  at the Fermi level in one-particle DOS 
$\rho_f(\omega)$. 

Even though a direct interpretation of the increase in terms of a
Fermi liquid picture (where the charge susceptibility is proportional 
to the DOS at the Fermi level) is not applicable since the Fermi liquid is 
formed only at very low temperatures, it still provides an intuitive way of 
understanding:
The increasing spectral weight at the Fermi level leads to an enlarged 
phase space volume for two-particle excitations and the charge susceptibility
is at least roughly proportional to the DOS at the Fermi level. This is supported by the fact, 
that $\chi_{charge}(0)$ increases logarithmically with decreasing 
temperature, which is also the case for $\rho_f(0)$.
But how strong the increase actually is and how it is influenced by the value of the Coulomb
repulsion $U$ cannot be deduced from the simplified  
Fermi liquid analogy. This rather depends on the details of the two-particle
correlations.

In the symmetric situation ($U=-2\e^f=2$) the charge susceptibility
increases only moderately and approaches the exact $T=0$ limiting 
value known from the Bethe ansatz,
which is indicated by the arrow at the left border of the 
Figure~\ref{fig:app-nca_staticCharge}.
For the asymmetric cases, the increase is considerably more pronounced.
Especially, the drastic low temperature increase for $U=4$
is rather unexpected. The absolute value of the susceptibility is 
even larger than  for the smaller values of $U$, which
is  counterintuitive  since  charge fluctuations should be
suppressed for larger $U$. 
However, the tendency
that for a given level position $\e^f$
the charge susceptibility in the asymmetric situation can increase with growing $U$  
is known from 
perturbation theory\cite{horvaticIVKondoPerturb85}  as 
a characteristic feature of valence fluctuation physics.
Valence fluctuations being at the origin of this enhanced low
temperature increase of the charge susceptibility are in agreement with the observation already made above,
that for $U=4$ the doubly occupied ionic orbital is outside the conduction 
band and the system is therefore  from the outset closer to the intermediate 
valence fixed point. 

Reference calculations with the NRG (not shown) 
indeed display the  characteristic features  of the charge susceptibility 
as shown in Figure \ref{fig:app-nca_staticCharge}: A maximum 
for temperatures of the order of the ionic level positions
$|\e^f|$ and $\e^f+U$ and an increase towards lower temperatures. 
In situations close to the valence fluctuation regime this increase
leads to an enhancement of the charge susceptibility by a factor
of about 10. However, the parameter values chosen for the 
$U=4$ ENCA-curve are 
not very close to the valence fluctuation fixed point. This  
is also reflected in  the magnetic susceptibility for $U=4$, which does not
show any signatures of the valence fluctuation regime, but rather exhibits 
behavior characteristic for the transition from a local moment to the 
strong coupling fixed point (not shown).
NRG calculations with parameter values similar to the ones 
chosen in this study, did show a low temperature increase,
but not as strong as observed with the ENCA. 

Altogether it can be concluded, that the ENCA does describe 
the charge fluctuations qualitatively right, but 
overestimates the influence of intermediate valence phenomena
at very low temperatures in the asymmetric case.    
\begin{figure}
  {  \scriptsize
    \includegraphics[scale=0.65]{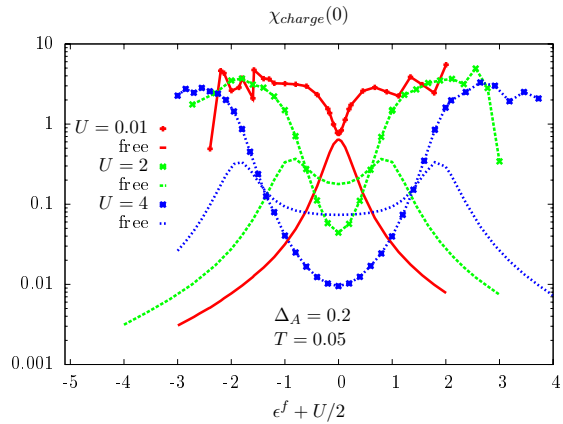}
  }
  \caption{Static charge susceptibility
    for a fixed $T=0.05$ and $\Delta_A=0.2$ as functions of the ionic level position $\e^f$ relative
    to the half-filling value $-U/2$ for various values of $U$.
    The conduction band was chosen to be constant with a half bandwidth of $W=10$.
    Curves without dots (``free'')
    are calculated without two-particle interactions, i.e.\ with  the particle-hole propagator
    of equation (\ref{eq:PHprop}).
  }
  \label{fig:app-nca_staticChiAsymCha}
\end{figure}

To make the range of applicability of the ENCA more clear, it is 
instructive to consider the charge susceptibility for fixed values of $U$ and 
$T$, varying the ionic level positions $\e^f$, which is shown in 
Figure \ref{fig:app-nca_staticChiAsymCha}. 
The particle-hole propagators already displayed in Figure \ref{fig:app-nca_staticChiSymMag}
are included as well (lines without dots).
The ENCA charge susceptibilities are always minimal for 
half filling ($\e^f+U/2=0$)  and 
 increases away from the symmetric case.
The absolute value of the charge susceptibility in the symmetric situation 
is  drastically reduced compared to the corresponding particle-hole
propagator for large values $U=2$ and $U=4$,
which indicates, that the two-particle correlations strongly suppress
charge fluctuations. In that situation, the susceptibility cannot
accurately be
described by the one-particle DOS alone and  independent though strongly renormalized
quasiparticles. 

On the logarithmic scale, the increase with growing distance from zero
can nicely be fitted with a parabola centered at zero, which corresponds 
to an exponential increase of the susceptibility,
$\chi_{charge}\sim e^{\alpha (\e^f+U/2)^2}$, $(\alpha>0)$.
This shows the strong influence of the asymmetry and
the contribution of valence fluctuations to the charge fluctuations.

However, the ENCA clearly fails for large asymmetries as the susceptibility  
saturates for $|\e^f+\frac{1}{2}U|>\frac{1}{2}U$\footnote{Additionally, the numerics
become unstable in these situations as it can be guessed from 
the strong fluctuations.}. 
In contrast, $\chi_{charge}(0)$  should decrease 
again (cf.\ \citet{kawakamiAsymSIAM82})
and approach the particle-hole propagator, due to the effective 
non-interacting nature. 
This is most drastic for the almost non-interacting case with $U=0.01$,
where, apart from reproducing the value at half filling quite accurately,
the curve goes the opposite direction as expected. 
The  values at which the downturn in the susceptibility should occur
correspond to  situations, where  both Hubbard  peaks in the 
one-particle spectrum (very roughly at $\e^f$ and $\e^f+U$) are either 
below or above the Fermi level,
corresponding to the empty- and fully occupied impurity
regimes. 

The ENCA is designed to describe
spin flip scattering and the magnetic exchange coupling correctly
but  it does not fully capture the physics of charge fluctuations outside the
Kondo regime. In situations, where the
unperturbed ground state is either the  empty
or doubly occupied ionic state, crossing diagrams neglected in the ENCA 
are vital to describe charge fluctuations accurately.
On the other hand,  magnetic fluctuations are still described
very  accurately in these situations (see Figures 
\ref{fig:app-nca_staticChiMomExact} and \ref{fig:app-nca_staticChiSymMag}).

\subsection{\label{sec:app-Dynsus} Dynamic susceptibilities}

\subsubsection{\label{sec:dynMag} Magnetic susceptibility}
\begin{figure}
  {  \scriptsize
    \includegraphics[scale=0.65]{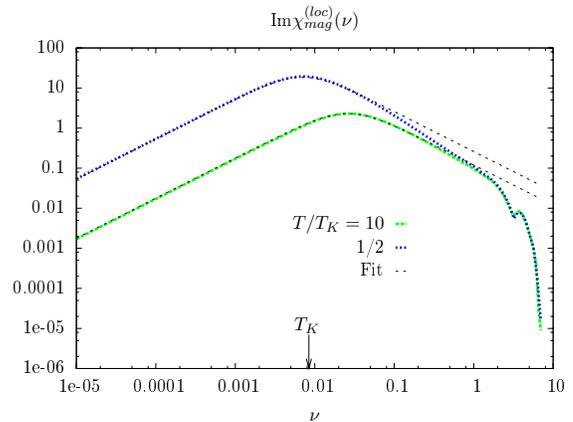}
  }
  \caption{The imaginary part of the dynamic magnetic susceptibility 
    for  $\e^f=-1$ two values of $U$ ($U=2,4$) 
    and two characteristic temperatures ($T=10T_K,T_K/2$) 
    in a  double-logarithmic plot. The corresponding 
    Kondo temperatures are indicated as arrows on the frequency axis.   
    All curves are calculated for $\Delta_A=0.3$ and a 3d-SC band DOS ($W=3$). 
   }
  \label{fig:app-nca_dynChiMagV472U2}
\end{figure}
The imaginary part of the dynamic magnetic susceptibility is shown in
Figure~\ref{fig:app-nca_dynChiMagV472U2}
for two different values of $U$ and two characteristic temperatures.  
The spectrum of the  susceptibility shows a pronounced maximum,
which is shifted to lower frequencies and increases considerably in height 
as the temperature is lowered. 
For temperatures below the Kondo temperature, the position   
of the maximum remains fixed at a value of the order of the Kondo temperature.

Also shown in the figure  are  fits with  a Lorentzian form  
\begin{align}
  \label{eq:app-ncaDynSusFit}
  \chi_{mag}^{fit}(\nu)&=\frac{\chi_0}{1-i\nu/\Gamma} \qquad,\quad\nu\in\mathbb{R}
  \quad,
\end{align}
which describe the low frequency susceptibilities very well.
The form (\ref{eq:app-ncaDynSusFit}) 
corresponds to an exponential spin relaxation with 
relaxation time $1/\Gamma$.
The line-width  $\Gamma$ is directly proportional 
to the NMR impurity nuclear spin-lattice relaxation 
rate $\Gamma\sim T_1$.\cite{shibaKorringaRelationSIAM75}

The relaxation rates $\Gamma$ extracted from susceptibilities
for various parameters follows a $\sqrt{T}$-law\cite{schmittPhD08} for high 
temperatures and saturates at a value of the order of $T_K$ at 
temperatures below $T_K$ (not shown), in accord with what 
was already found earlier.\cite{bickers:nca87b,jarrell:dynamicSusSIAM91,anders:PostNCA95}

The  physical picture behind these findings is quite clear:
Upon lowering the temperature, the local moment of the impurity
becomes increasingly coupled to the surrounding spin of the band 
electrons resulting in an enhanced response.   
At temperatures of the order of or lower than the Kondo temperature,
the local  Fermi liquid state is approached in which the local spin is screened and
a local Kondo singlet is formed with a ``binding energy'' of about $T_K$. 
Therefore the maximum in the spin excitations spectrum, as well as the NMR 
relaxation rate, both are pinned  at an energy of the order of $T_K$.

\begin{figure}
  {  \scriptsize
    \includegraphics[scale=0.65]{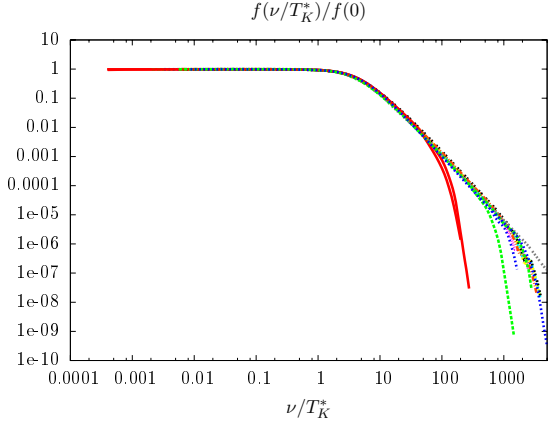}
  }
  \caption{ Scaling function $f(\nu/T_K^*)/f(0)$ from equation~\eqref{eq:app-ncaScaling}
    for the following parameter sets: 
    ($\e^f\!=\!-0.5; U\!=\!1; T/T_K\!=\!10,1$), ($\e^f\!=\!-1; U\!=\!2; T/T_K\!=\!1$), 
    ($\e^f\!=\!-1.5; U\!=\!3; T/T_K\!=\!1$), ($\e^f\!=\!-2; U\!=\!4; T/T_K\!=\!1$), 
    ($\e^f\!=\!-1; U\!=\!3; T/T_K\!=\!1$), ($\e^f\!=\!-1; U\!=\!4; T/T_K\!=\!1,1/2,1/3,1/5,1/7,1/10$) 
    and  ($\e^f\!=\!-1; U\!=\!8; T/T_K\!=\!1$).
    All curves are calculated for $\Delta_A=0.3$ and a 3d-SC band DOS. 
   }
  \label{fig:app-nca_dynChiMagScale}
\end{figure}

\citet{jarrell:dynamicSusSIAM91} also found that the function
\begin{align}
  \label{eq:app-ncaScaling}
  f(\nu)&=\frac{\pi T_K}{2\chi_{mag}(0)}\frac{\mathrm{Im}\chi_{mag}(\nu)}{\nu}
\end{align}
shows universality and depends only on $T/T_K$. 
Figure~\ref{fig:app-nca_dynChiMagScale} shows this function normalized to its 
zero frequency value for various parameter sets. All graphs can be collapsed 
onto one single curve showing the universal shape of the function $f$
for low energies. 

In order to achieve scaling a guess for the actual 
Kondo temperature $T_K^*$ has to be used. In contrast, the  $T_K$-value
calculated by equation~\eqref{eq:app-nca_TKENCA} and used in this work
does not represent the exact physical low energy scale  $T_K^*$, but only
provides an order of magnitude estimate. In the universal regime
with a flat $c$-band DOS this should not make any difference, but since 
we are using the 3d-SC DOS, non-universal corrections enter 
for different $\e^f$ and $U$.
The value of the ``real'' 
Kondo temperature could have been extracted from fits of
the calculated susceptibilities to the universal curve of the
susceptibility as described by \citet{jarrellTransportSIAM91}.

The rapid decrease of the  curves  in the figure for frequencies 
of $\nu/T_K^*\gtrapprox 100$ also stems from the 
finite bandwidth of the 3d-SC conduction band used for these calculations.

The above findings clearly confirm, that the dynamics
of the impurity spin is solely determined by the antiferromagnetic exchange
between the impurity- and conduction electron  spins.
Even in the asymmetric situation, the only relevant energy scale for 
magnetic fluctuations  of the impurity is the Kondo temperature $T_K$ at 
low temperatures.

\subsubsection{\label{sec:dynCha} Charge susceptibility}

\begin{figure}
  {  \scriptsize
    \includegraphics[scale=0.65]{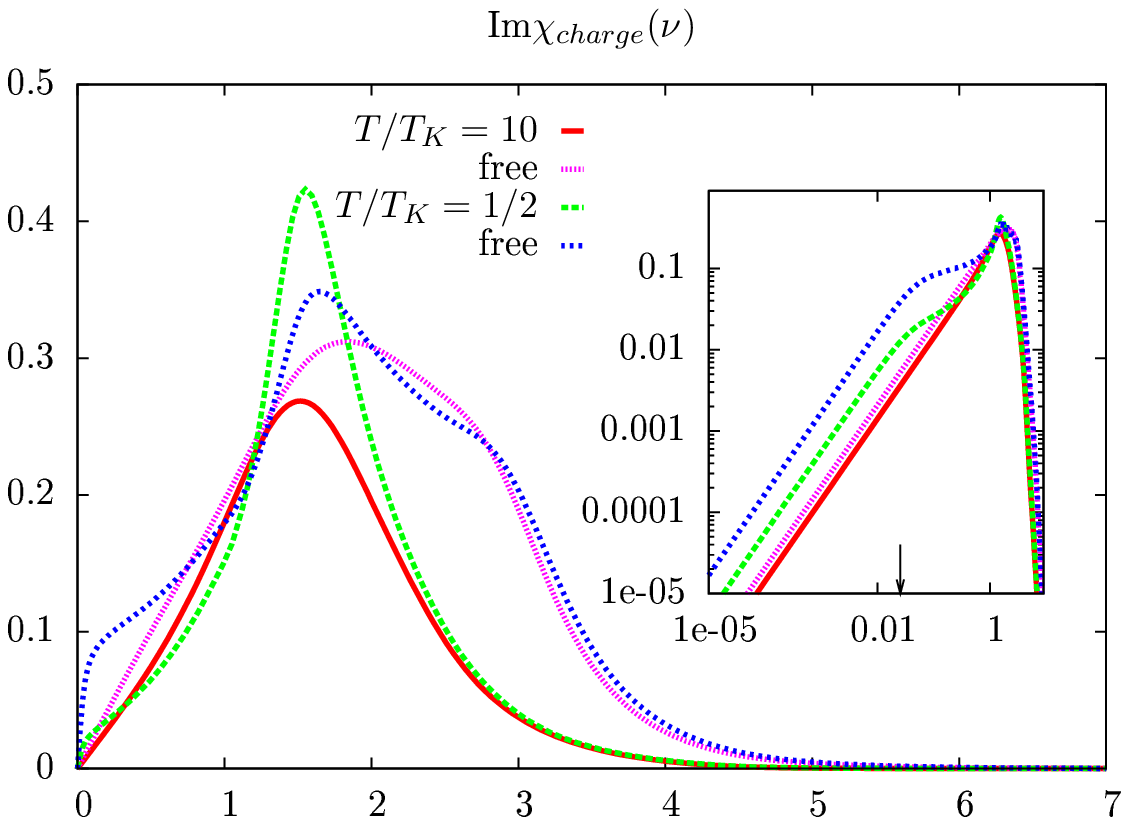}\\
    \includegraphics[scale=0.65]{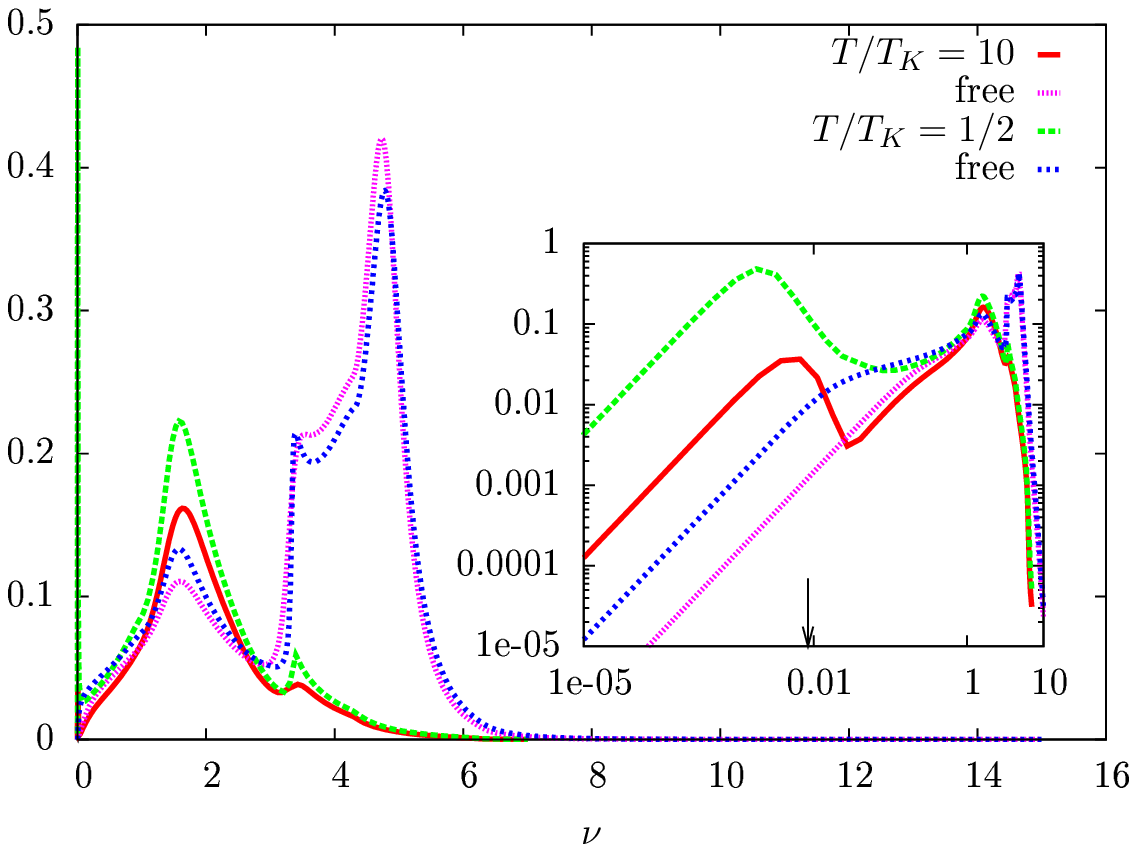}
  } 
  \caption{Imaginary part of the dynamic charge 
    susceptibility for the symmetric  ($\e^f=-1$, $U=2$, upper) and  
    asymmetric ($\e^f=-1$, $U=4$, lower) case for
    two characteristic  temperatures $T=10T_K$ and $T=T_K/2$.
    The curves for the particle-hole propagators (``free'') are shown as well.
    The insets show the corresponding  low energy region on a 
    double-logarithmic scale.
    The Kondo temperature for each case is indicated by the red arrow 
    on the frequency axis in the low energy insets.  
    The calculations are done for a 3d-SC band and
    $\Delta_A=0.3$.
  }
  \label{fig:app-nca_dynChiChargeV578}
\end{figure}

The  imaginary part of the dynamic charge susceptibilities for two 
different Coulomb repulsions and 
characteristic temperatures, calculated with a 3d-SC band DOS,
are shown in Figure~\ref{fig:app-nca_dynChiChargeV578}. 
In the spectra the characteristic features stemming from excitations
involving the Hubbard peaks  at energies 
around $|\e^f|$ and $\e^f+U$  are clearly visible. 

In  the symmetric case ($U=2$), the height of the peak at $\nu\approx |\e^f|$ 
is about twice of the one in the asymmetric case ($U=4$). 
This is due to the doubled phase-space volume for the symmetric
situation with excitation 
energies matching, 
$\e^f+U=|\e^f|$, while
for $U=4$ the upper Hubbard peak is moved 
to higher energies. 

Also shown in the graphs are the local particle-hole propagators
of equation~(\ref{eq:PHprop}) (labeled as ``free'').
These show characteristic features of the Hubbard peaks too, but 
the most prominent difference to the 
fully interacting susceptibility  
is the strong suppression of the  high energy response in the latter.
For example, the broad excitation continuum in the particle-hole propagator
of the asymmetric case ($U=4$)
for energies  in the range $3\leq\nu\leq 6$
is reduced to a very small peak at $\nu\approx 3$ in the fully interacting
susceptibility.

$P^f(\nu)$ is just a measure for the 
phase space volume for statistical independent particle-hole
excitations, which are described by the one-particle DOS $\rho_f$.
The quasiparticles
and -holes at the Fermi level, on the other hand, are strongly
correlated, leading to an effective suppression of the available phase
space volume.

The role of the low energy quasiparticles can be studied by
comparing the particle-hole propagator and the interacting
susceptibilities at low energies for the symmetric situation
($U=2$, inset).
The response via the particle-hole propagator for $\nu\leq T_K$ shows an
increase for lower temperatures, which stems from the quasiparticle-quasihole
excitations within the growing Kondo resonance. In the fully interacting 
susceptibility this increase is approximately an order of magnitude
smaller, clearly showing the effect of correlations in the two-particle response.  

Surprisingly, for the asymmetric case ($U=4$) this trend 
is reversed and 
the interacting susceptibility is enhanced over the particle-hole 
propagator for excitation energies 
smaller than the Kondo scale $\nu\leq T_K$ (see inset).
This very pronounced low energy response is produced 
by quasiparticle-quasihole excitations 
in the local Fermi liquid phase at low temperatures.
The fact that these excitations are strongly enhanced in the 
asymmetric  case compared  to the symmetric situation is associated 
with the presence of strong valence fluctuations, as was already discussed for the static
susceptibility above. 

Even though such a low energy enhancement was not reported in NRG 
calculations for the charge susceptibility\cite{sakaiSIAMRG89,frota:ChargeSusSIAMnrg91} 
an inspection of the low energy part of the many-body spectrum 
obtained with the NRG~\cite{andersNRGLecture04} 
and preliminary NRG-calculations indeed suggest the possibility 
of an enhanced charge response for  asymmetric situations.

A similar but not as strong temperature dependent increase in the dynamic 
susceptibility  for temperatures of the order of the
Kondo temperature was already found for larger orbital degeneracy and 
$U=\infty$ with the NCA~.\cite{brunner:ChargeSusSIAM97}

Therefore we argue, that these findings for the dynamic charge 
susceptibility are in accord with the ones discussed in the previous section
for the static charge susceptibility. The observed increase of the charge 
response for energies smaller than the Kondo scale is indeed 
physical and due to the influence of the valence fluctuations
of the low temperature Fermi liquid. However, the magnitude
of the enhancement shown in Figure~\ref{fig:app-nca_dynChiChargeV578}
is arguable, especially for the choice of parameters in the present
calculation.

\section{\label{sec:concl} Conclusions}

We have studied the SIAM within a conserving 
approximation, the ENCA, for a variety of model parameters. 
It was shown, that the ENCA constitutes a very accurate
approximation for the static and dynamic  one- and two-particle
quantities of that model for temperatures down to a fraction of the 
Kondo temperature. 
It considerably improves the Fermi liquid properties
and cures  shortcomings of the (S)NCA, like the
removal of the divergence of
$\lim_{\nu\to 0}\frac{\mathrm{Im}\chi_{mag}(\nu)}{\nu}$ at zero
temperature.

In symmetric situations ($2\e^f+U=0$), the static magnetic and charge
susceptibilities were  shown to be in excellent agreement with the exact Bethe ansatz
results. This was even true for cases with very small Coulomb
interaction $U$, which could not be expected from the beginning, since approximations 
within direct perturbation theory with respect to the hybridization  
usually have problems describing the non-interacting case. 

The static magnetic susceptibility is in excellent agreement with exact Bethe ansatz 
results in the asymmetric situation. This holds also in cases with strong valence 
fluctuations, such as for $\e^f=0$ or in the empty-  and fully occupied 
orbital regimes.

However, the static charge susceptibility in the asymmetric model
is described accurately only in situations, where the singly 
occupied impurity valence state
represents the unperturbed ($\Delta_A=0$) ground state. 
In addition, even though we believe that the qualitative features of the
charge susceptibility in asymmetric situations are captured by the presented
calculations, the influence of valence fluctuations is probably 
overestimated for too low temperatures. 
This confirms the expectation, that crossing diagrams, which are 
neglected in the ENCA, are essential for the quantitative description 
of situations with strong valence fluctuations, where the impurity 
occupation is statistically fluctuating.  
This also is  in accord with the known pathologies in the one-particle 
spectral function. There,
charge and magnetic fluctuations both contribute and the overestimation
of the charge excitations at very low temperatures leads to 
the overshooting of the Kondo resonance and the observes spikes in
the  DOS.

The dynamic magnetic susceptibility is dominated by Kondo screening
of the impurity spin. The ENCA  correctly reproduces the
temperature and other parameter dependencies of the magnetic excitations,
and also the scaling found in previous studies is obtained. 

The dynamical charge spectrum shows a severe suppression of 
high energy excitations due to correlations, when compared to
the particle-hole propagator, which would represent the 
susceptibility of independent renormalized quasiparticles.
Additionally the low energy response for excitation energies smaller than the 
Kondo temperature is also strongly suppressed in the symmetric case,
due to the same correlations between low energy quasiparticles.  

In the asymmetric situation the low energy charge response 
is drastically enhanced and an additional peak emerges. This
enhancement is attributed to the presence of the valence fluctuation
fixed point in the asymmetric model. Such an enhancement seems quite
probable so that only the steepness of the increase calculated within
the ENCA for parameter values chosen is arguable.

With these findings, the prospects of describing 
two-particle dynamics of lattice systems within the DMFT are very promising and
results will be presented in a subsequent publication.\cite{inprep,schmittPhD08}

\begin{acknowledgments}
  The authors acknowledge fruitful discussions with E.\ Jakobi and F.B.\ Anders. 
  One of us (SS) especially thanks F.B.\ Anders  for providing him with
  his NRG code to perform the reference calculations mentioned in the text,
  and acknowledges support from the DFG under Grant No.\ AN 275/6-1.
\end{acknowledgments}




\end{document}